\documentclass[preprint2]{aastex}

\shorttitle{Exact Solution to Finite Temperature SFDM}
\shortauthors{V.H. Robles & T. Matos}

\begin{document}


\title{Exact Solution to Finite Temperature SFDM: Natural Cores without Feedback}


\author{Victor H. Robles\altaffilmark{*} and T. Matos\altaffilmark{**}}
\affil{Departamento de F\'isica,Centro de Investigaci\'on y de Estudios Avanzados del IPN, AP 14-740, 0700  D.F., M\'exico}


\altaffiltext{*}{Electronic address: \tt{vrobles@fis.cinvestav.mx}}
\altaffiltext{**}{Electronic address: \tt{tmatos@fis.cinvestav.mx}}


\begin{abstract}
Recent high-quality observations of low surface brightness (LSB) galaxies have shown that their dark matter (DM) halos 
prefer flat central density profiles. On the other hand, the standard cold dark matter model simulations predict a more cuspy behavior. 
One mechanism to reconcile the simulations with the observed data is
the feedback from star formation, this might be successful in isolated dwarf galaxies but its success in LSB galaxies 
remains unclear.
Additionally, including too much feedback in the simulations is a double-edged sword, in order to obtain a cored DM distribution 
from an initially cuspy one, the feedback recipes usually require to remove a large quantity of baryons from the center of galaxies, 
however, some feedback recipes produce twice more satellite galaxies of a given luminosity and with much smaller
mass to light ratios from those that are observed.
Therefore, one DM profile that produces cores naturally and that does not require large amounts of feedback would be preferable.
We find both requirements to be satisfied in the scalar field dark matter model. Here, we consider that the dark matter is an 
auto-interacting real scalar field in a thermal bath at temperature T with an initial $Z_2$ symmetric potential, 
as the universe expands, the temperature drops so that the $Z_2$ symmetry is spontaneously broken and the field rolls down to a new
minimum. We give an exact analytic solution to the Newtonian limit of this system and show that it can satisfy the 
two desired requirements and that the rotation curve profile is not longer universal.

\end{abstract}


\keywords{galaxies:formation--galaxies:halos ---galaxies:individual (NGC 1003,
NGC 1560, NGC 6946)--galaxies:fundamental parameters }



\section{INTRODUCTION}

The longstanding core/cusp discussion, whether the central dark matter (DM) profiles in dwarfs and low
surface brightness (LSB) galaxies are more core-like and rounder than the standard cold dark matter (CDM) model predicts, 
remains an open issue (\citet[]{eym09,blo10} for a recent review). 
So far, the core profiles most frequently used in the literature and that best fit the observations are 
empirical \cite[]{bur95,kuz10}. Though they are useful to characterize properties of galaxies, it is necessary to 
find a theoretical framework which naturally produces the cores, especially since more and more recent high-quality observations of 
LSB galaxies suggest that the core-like behavior ($\rho \sim r^{-0.2}$) is preferred in the central regions of dwarf 
and LSB galaxies \cite[]{seheon11,rob12,blo01}. 
This is a problem to the CDM model which prefers $\rho \sim r^{-1}$ at small $r$ \citep{nav10}.
Though the latest simulations can reach $\rho \sim r^{-0.8}$ \citep{nav10, mer06,gra06}, which is not in total agreement 
with observations.
The current trend to solve the core/cups discrepancy in the CDM model is to include the dynamics of the baryonic 
component into DM simulations \citep{gov10,gov12,mac11,sti11,rom08}. By including feedback from star formation in 
simulations of field dwarf galaxies, \cite{gov12} managed to change an initially cuspy DM halo into a core-like halo. 
However, it still remains to be seen if the same feedback recipes used in dwarf galaxies work 
as well in massive LSBs. As pointed out in \citet{kuz11} this seems unlikely, it is
necessary to show that there is an accord between the surface gas density in LSBs and the amount of it obtained in 
CDM simulations \citep{bro12,sti11,guo10,mor11,kuz11b}. We know LSBs are a large portion of the total galaxies that are observed
\citep{mcg95}, therefore, as far as there is not yet an agreement with LSBs and CDM simulations 
it worths exploring alternative models.

As another problem, \cite{saw12} find in their simulations twice as many satellites of a given luminosity around a Milky Way size host 
halo and found that, if galaxies are intrinsically cored at infall then the transition from high gas mass irregulars to dwarf spheroidals 
cannot be only due to tidal stripping. They found this by considering, in addition to feedback, the effect that surrounding galaxies 
have on its host halo by means of tidal interation. 
They find that the DM halos of satellites are more strongly affected than their stellar component by tidal interactions, 
and conclude from their simulations that it is difficult to reconcile the observed high total mass-to-light ratios in dwarf 
spheroidal galaxies (dSphs) with those found on their simulated counterparts under the CDM paradigm. 
A solution to this discrepancy should also be addressed in any alternative DM model.

There are several models in the literature that are addressing these and some other problems
\citep{avi01,cem05,str07,spe00,mag12a}.
Models that slightly modify CDM, like the warm dark matter (WDM) and self-interacting dark matter (SIDM) models, 
haven't been able to solve these discrepancies yet \citep{nav10,kuz11,zav09,dav01,yos00}.
There are models that modify gravity like f(R) theories \citep{def10} and MOND 
\citep{mil10,san09}, but they are currently more at the effective-theory level rather than at a fundamental one. Nevertheless,
there are some rotation curves fits of LSBs galaxies in MOND models whose fits are almost perfect to the observed data, for instance 
NGC1560 \citep{san09}.

One model that has received much attention is the scalar field dark matter (SFDM) model.
It is our aim to show that in this model there is an scenario of galaxy 
formation (described in section 2) different from the standard model used in CDM simulations and that naturally produces 
core density profiles, reproduces rotation curves of large and small galaxies on equal footing as MOND and empirical dark matter 
models do, but that may not need of large amounts of feedback to agree with observations. 

In previous works it has been verified that the SFDM model reproduces cosmological observations as well as 
CDM \citep{rod10,sua11,mag12a,har11}.
In cosmological scales, choosing a scalar field mass $m\sim$10$^{-22}$eV/$c^2$ gives the same cosmological density 
evolution that CDM\citep{mat09,cha11} and is consistent with the acoustic peaks of the cosmic microwave 
background radiation\citep{rod10}. Moreover, in \cite{mat01} and \cite{hu00} they found that it suffices that $m<$10$^{-17}$eV/$c^2$ for 
the SF to condensate. On the small scale regime, \cite{rin11} studied the formation of vortex in SFDM halos 
and found constraints on the boson mass in agreement with previous works\citep{kai10,zin11}. \cite{lor12} 
analysed the dynamics of Ursa Minor and its stellar clump within the SFDM paradigm and found a good agreement with observations 
with $m\sim$10$^{-22}$eV/$c^2$ . In the galactic scale the self-interaction plays an important role, however, its constraints 
are still not very tight, especially since the interaction parameter always appears entangled with $m$ when fitting 
rotation curves\citep{rob12,boh07}. The existence of two parameters allows us to use a smaller value of the 
scalar field mass \citep{boh07,har11}, nevertheless, $m\sim$10$^{-22}$eV/$c^2$ is also possible, therefore, 
there is no incompatibility with cosmological and galactic scales parameters.

The article is organized as follows, in section 2 we describe the SFDM model to be analysed in this paper, 
in section 3 we give our results and section 4 is devoted to conclusions.  
 
\section{SFDM MODEL}


\subsection{Previous work and unsolved issues}

The idea was first considered by \citet{sin94} and independently introduced by \citet{guz00}.
In the SFDM model the main hypothesis is that the dark matter is an auto-interacting real scalar field 
that condensates forming Bose-Einstein Condensate (BEC) ``drops'' \citep{mag12a}. We interpret these BEC drops as the halos
of galaxies, such that its wave properties and the Heisenberg uncertainty principle stop the DM phase-space density from 
growing indefinitely, and thus, it avoids cuspy halos and reduces the number of small satellites \citep{hu00}.

In the SFDM model, the scenario of galaxy formation proposes that galactic halos form by condensation of a scalar field (SF) 
with an ultra-light mass of the order of $m \sim 10^{-22}$eV ( units where the speed of light $c$=1). 
From this mass it follows that the critical temperature of condensation of the scalar field is T$_\mathrm{crit}\sim m^{-5/3}\sim$TeV, 
which is very high, thus, BEC drops are formed very early in the universe. 
It has been proposed that these drops are the halos of galaxies \citep{mat01}, i.e., that halos are gigantic
clumps of SF. 

The galaxy size DM halos can be described in the non-relativistic regime, where they can 
be seen as a Newtonian gas. When the SF has self-interaction, we need to add a quartic term 
to the SF potential and in the Newtonian limit the equation of state of the SF is that of a polytrope of index 1 \citep{sua11,har11}. 
Some studies of the stability of these SF configurations have shown that stable large scale configuration are not prefered 
\citep{col86,bal98,val11}, though the critical mass for stability depends of which paramenter values were 
used, all reach the same conclusion, very large configurations, like those of cluster scales (masses of M$\geqslant$10$^{13}M_{\odot}$),
are usually unstable, therefore, these structures were most likely form just as in the CDM model, by hierarchy \citep{mat01,sua11},
 i.e., by mergers of smaller halos.The idea of the SFDM model is the following, after inflation big structures start hierarchically to grow up like in the CDM model and its growth will 
be boosted by the SB mechanism. Inside of them, like galactic size halos will be formed by condensation.
Thus, all predictions of the CDM model at big scales are reproduced by 
SFDM \citep{col86,gle88,mat01,cha11}.
Although the model is notably succesfull, there are at least two reasons to complement 
the model in addition to finding an explanation to the two discrepancies discussed in the introduction. 

The first one is found when we consider the fully condensed system at temperature T=0, the fits to rotation curves (RCs) of 
LSBs \citep{rob12,boh07} show deviations from the observed data at large radii because 
DM halos are modeled only by considering the ground state (complete condensation), whose 
DM density and velocity profile are given by \citep{boh07} 
\begin{mathletters}
\begin{eqnarray} 
\rho^0(r)  & = & \rho^0_{0} \frac{\sin( \pi r/ \hat R)}{\pi r/ \hat R} \label{zerotden},  \\ 
V^2_{0}(r)   & = & \frac{4 \pi G \rho^0_{0}}{K^{2}} \biggl( \frac{\sin (Kr)}{Kr} - \cos(Kr) \biggr ), \label{zerotvel}
\end{eqnarray}
\end{mathletters}
where $K= \pi/ \hat R$, $\rho^0_{0} = \rho^0(0)$ is the central density, and the halo radius determined by $\rho(\hat R)=0$ is
\begin{equation}
 \hat R= \pi \sqrt{\frac{\hbar^{2} b}{G m^{3}}}, 
\end{equation}
here $\hbar$ is Planck's constant divided by $2 \pi$, $m$ is the mass of the DM particle, G is the gravitational constant, and $b$ is
the scattering length. The latter is related to the coupling constant $\lambda'$ by $\lambda'=4 \pi \hbar^2 b/m$.
The density profile depends on two fitting parameters, the central density and a length scale $\hat R$. 
However, $\hat R$ depends only on fixed parameters, the interaction parameter $\lambda'$ and the mass of the SF particle, 
which implies that it should not vary from galaxy to galaxy. However, when fitting rotation curves of galaxies, as $\hat R$ is 
a fitting parameter it gets different values for each galaxy \citep{rob12,boh07}, this is something contradictory and 
represents a problem for the simplest SFDM model. 

The second problem lies in the fast decrease of the velocity profile (equation (\ref{zerotvel})) after its maximum value,
such decrease causes a disagreement between the fits and the observed data in large galaxies because the data usually remain
``flat'' until the outermost regions \citep{swa00}. In addition to this, and for large galaxies, when we aim for the
best fit to the velocity maximum, we obtain a worse fit in the outer regions, i.e. the better the fit to the velocity 
maximum in the RC the worse the fit becomes in the outer regions and viceversa.

One approach to solve the problems was consider by \cite{mat07} and later by \cite{har11b}. It consisted of including 
finite temperature of the DM in the DM halos. \cite{har11b} show that by including a small 
correction to the pure condensed state, albeit in a different way from the one in this paper, it is possible to solve the 
first and partially solve the second discrepancy. 
The inclusion of temperature T, does mainly two things: 1) it makes the halo radius temperature-dependent, thus it's not 
fixed for all galaxies anymore and 2) it lifts the RC fit in the outer region and keeps it flat until the last value. 
However, consequence 2) is negligible when the halo temperature is T$<$0.5T$_\mathrm{BEC}$, where T$_\mathrm{BEC}$ is the 
critical temperature of Bose Einstein condensation, if T$_\mathrm{BEC}\sim$TeV ($c$=1), we would expect that present halos
be well approximated by equation (\ref{zerotvel}) and hence, we will be unable to simultaneously obtained a good fit to the 
RC maximum and the RC outer regions.
Though this last problem is not readily visible in galaxies with small radius (outer radius of $\leq$10 kpc), it is conspicuous 
in large galaxies, therefore problem 2) is just partially solved.

Thus, an alternative approach to solve these two SFDM discrepancies considers non-condensed SF configurations at T=0, i.e.
SF configurations in exited states \citep{bal98,ure10,ber10}. These configurations fit RCs up to 
the last data point and can even reproduce the wiggles seen at large radii in high-resolution observations 
\citep{sin94,col86}. The problem found in this kind of solution was that the configurations 
required for good fits (4-5 exited states) were unstable and decayed to the ground 
state in a short time \citep{sid03}. Therefore, we will expect to see today that most DM halos 
are in their ground state, which means that the disagreement at large radii would remain.

\subsection{Finite temperature SF scenario}

Motivated to solve all these issues we consider the following scenario which includes temperature of the DM 
and the exited states of the SF.

The idea is that the dark matter is a spin-0 scalar field $\Phi$, with a repulsive interaction embedded in
a thermal bath of temperature T, we also consider the finite temperature corrrections up to one-loop in the perturbations.
This system is described by the potential \citep{kol87,dal99} 
\begin{equation}
V (\Phi) = -\frac{1}{2} \frac{\hat m^2 c^2}{\hbar^2} \Phi^2+\frac{\hat\lambda}{4}\Phi^4 + \frac{\hat\lambda}{8}k_\mathrm{B}^2T^2\Phi^2 
-\frac{\pi^2 k_\mathrm{B}^4T^4}{90 \hbar^2 c^2}. 
\label{eq:pot}
\end{equation}
for the case when $k_\mathrm{B}$T$>>$ $\hat m c^2$. Here $k_\mathrm{B}$ is Boltzmann's constant, $\hat \lambda = \lambda/(\hbar^2 c^2 )$ is 
the parameter describing the interaction, $\hat \mu^2$:= $\hat m^2 c^2$/$\hbar^2$ is a parameter, and T is the temperature of the thermal bath.
The first term in $V(\Phi)$ relates to the mass term, the second to the repulsive self-interaction, the third to the 
interaction of the field with the thermal bath, and the last to the thermal bath only.

At some high enough temperature in the early universe the SF interacts with the rest of the matter and due to the expansion 
of the universe, its temperature will keep decreasing. Eventually, when the temperature is sufficiently small, the SF decouples 
from the interaction with the rest of the matter and follows its own thermodynamic history, while the background keeps cooling 
down due to the expansion.
Moreover, as the temperature continues decreasing, the field will reach the minimum of the potential in $\Phi$$\approx 0$, this 
initial minimum of the potential eventually becomes a local maximum and after this moment, 
the initial $Z_2$ symmetry of the potential $V(\Phi)$ will be broken. 
The latter happens at a critical temperature T$_C$ given by 
 \begin{equation}
  k_\mathrm{B}T_C=\frac{2\hat mc^2}{\sqrt{\lambda}}.  \label{eq:Tc}
 \end{equation}
We will see that the critical temperature determines the moment in which the DM fluctuations can start growing, 
they do it from the moment when T$<$T$_C$ until they reach a stable equilibrium point, 
for example in $\Phi_\mathrm{min}^2$=$k^2_\mathrm{B}(T_C^2-T^2)/4$(see section 2.3).

\subsection{Evolution equations}

The perturbed system of a scalar field with a quartic repulsive interation but with temperature zero has been studied before 
\citep{col86,ure10}. Following the same procedure we study the evolution of the 
SF in a FRW universe. We write the metric tensor as $\mathbf{g}= \mathbf{g}^0 + \delta \mathbf{g}$, 
where $\mathbf{g}^0$ is the unperturbed FRW background metric and $\delta \mathbf{g}$ the perturbation. The perturbed line element in conformal time $\eta$, is (we take $c$=1 in 
this subsection)
\begin{eqnarray}
ds^2 = a(\eta)^2(-(1+2\psi)d\eta^2+2B,_id\eta dx^i \label{eq:metric1} \nonumber\\
  + a(\eta)^2[(1-2\phi)\delta_{ij}+2E,_{ij}]dx^idx^j,  
\end{eqnarray}
with $a$ the scale factor, $\psi$ the lapse function, $\phi$ gravitational potential, B the shift, and E the anisotropic potential.
We separate the energy-momemtum tensor and the field as $\mathbf{T}=\mathbf{T}_0+\delta \mathbf{T}$ and
$\Phi(x^{\mu})=\Phi_0(\eta)+\delta \Phi(x^{\mu})$ respectively.
As we are studying the linear regime $\delta \Phi(x^{\mu}) << \Phi_0(\eta)$, we can approximate
$V(\Phi) \approx  V(\Phi_0)$. We work in the Newtonian gauge where the metric tensor $\mathbf{g}$ becomes diagonal and 
as a result, in the trace of the Einstein's equations the scalar potentials $\psi$ and $\phi$ are identical, therefore, 
$\psi$ relates to the gravitational potential. 

Changing to the cosmological time $t$ using the relation $(d/d\eta)$=$ a(d/dt)$, the perturbed Einstein's equations 
$\delta G^i_j$= 8$\pi$G$\delta T^i_j$ to first order for a scalar field in the Newtonian gauge (E=0=B) are
\begin{mathletters}\label{eq:sfmet}
\begin{eqnarray}
-8 \pi G \delta\rho_{\Phi}&=&6H(\dot{\phi}+H\phi)-\frac{2}{a^2}\nabla^2\phi, \\
8\pi G \dot{\Phi}_0\delta\Phi,_i&=&2(\dot{\phi}+H\phi),_i , \\
8\pi G\delta p_{\Phi}&=&2[\ddot{\phi}+3H\dot{\phi}+(2\dot{H}+H^2)\phi] 
\end{eqnarray}
\end{mathletters}
with $\dot{}=\partial/\partial t$ and $H=(\ln a)\dot{}$. 
The perturbed density $\delta\rho_{\Phi}$ and the perturbed pressure $\delta p_{\Phi}$ are defined in terms of
the perturbed energy momentum tensor as
\begin{eqnarray}
\delta T^0_0&=&-\delta\rho_{\Phi}=-(\dot{\Phi}_0\dot{\delta\Phi}-\dot{\Phi}_0^2\psi+V,_{\Phi_{0}}\delta\Phi), \nonumber \\
\delta T^0_i&=&-\frac{1}{a}(\dot{\Phi}_0\delta\Phi,_i),  \nonumber \\
\delta T^i_j&=&\delta p_{\Phi}=(\dot{\Phi}_0\dot{\delta\Phi}-\dot{\Phi}_0^2\psi-V,_{\Phi_{0}}\delta\Phi)\delta^i_j. \label{eq:Tperturbado}
\end{eqnarray}
Systems (\ref{eq:sfmet}) and (\ref{eq:Tperturbado}) describe the evolution of the scalar perturbations.
To study the evolution of the SF perturbations we use the perturbed Klein-Gordon equation 
\begin{eqnarray} \label{kgl}
 \ddot{\delta\Phi}&+& 3 H \dot{\delta\Phi}-\frac{1}{a^{2}} \mathbf{\nabla}^2\delta\Phi+ V,_{\Phi_0\Phi_0}\delta \Phi + \nonumber \\
 &+&2V,_{\Phi_0}\phi-4\dot{\Phi}_0\dot{\phi}=0.   
\end{eqnarray}
Equation (\ref{kgl}) can be rewritten as:
 \begin{equation}
 \Box \delta\Phi+\left.\frac{d^2V}{d\Phi^2}\right|_{\Phi_0}\delta\Phi+\left.2\frac{dV}{d\Phi}\right|_{\Phi_0}\phi -4\dot{\Phi}_0\dot \phi=0,
 \label{eq:KG}
 \end{equation}
where the D'Alambertian operator is defined as
 \begin{equation}
  \Box := \frac{\partial^2}{\partial t^2}+3H\frac{\partial}{\partial t}-  \frac{1}{a^2} \mathbf{\nabla}^2\label{eq:Box}\nonumber.
 \end{equation}
Essentially equation (\ref{kgl}) represents a harmonic oscillator with a damping $3H\dot{\delta\Phi}$ and an extra 
force $-2\phi V,_{\Phi_0}$.  Equation (\ref{kgl}) has oscillating solutions if $(V_{,\Phi\Phi} -\frac{1}{a^{2}}\mathbf{\nabla}^2)\delta\Phi$ 
is positive. This equation contains growing solutions if this term is negative, that is, if $V_{,\Phi\Phi}$ is  negative enough. 
From here we see an important feature. These perturbations grow only if $V$ has a maximum, even if it is a local one. 
Here, the potential is unstable and during the time when the scalar field remains in the maximum, the scalar field fluctuations 
grow until they reach a new stable point. 
If we use equation (\ref{eq:pot}) the transition from a local minimum to a local maximum 
happens when T$=$T$_C$, thus, we see why T$_C$ determines the moment in which the DM fluctuations can start growing.
This implies that the galactic scale halos could have formed within this period and with similar features. 
Finally for the background field equation we have that $\phi$$\approx$0, therefore its equation reads
\begin{equation}
\ddot \Phi_0 + 3H \dot \Phi_0 + V,_\Phi (\Phi_0) =0. \label{backg}
\end{equation}
We have that $\Phi_0$ depends only on time.

\begin{figure}
\plotone{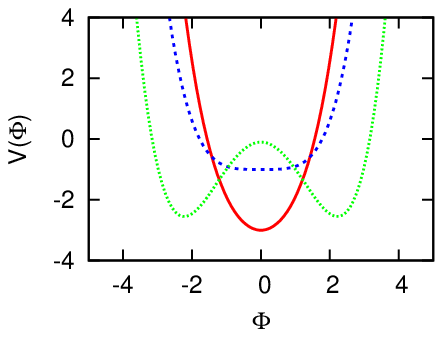}
\caption{We plot the qualitative behavior of $V(\Phi)$ vs $\Phi$ for three different values of T.
The solid line (red in the online version) is when T$>$T$_C$, here the system oscillates around the minimum and possesses a $Z_2$ symmetry, 
the dashed line (blue in the online version) is when the SB takes place, this happens when T=T$_C$, and the dotted line 
(green in the online version) is for T$<$T$_C$, in this period $\Phi$ can roll down to a new minimum and there is no longer
$Z_2$ symmetry.}
\label{fig1}
\end{figure}

We now suppose that the temperature is sufficiently small so that the interaction 
between the SF and the rest of matter has decoupled, after this moment the field stops interacting with the rest of the particles.
We also assume that the symmetry break (SB) took place in the radiation dominated era in a flat universe.
We mentioned that after the SB, the perturbations can grow until the reach their new minimum, thus, each perturbation 
has a temperature at which it forms and separates from the background field following its own evolution, we denote 
this temperature of formation by T$_\Phi$:=T$(t_\mathrm{form})$, $t_\mathrm{form}$ is the time in which the halo forms.
Under these assumptions the equation for an SF perturbation which is formed at T$_\Phi$ reads
\begin{eqnarray}
\Box \delta \Phi &+&\frac{\hat\lambda}{4 } [k_\mathrm{B}^2 (T^2_{\Phi}-T^2_C) + 12\Phi_0^2]\delta \Phi -4\dot{\Phi}_0\dot{\phi} \nonumber \\
&+&\frac{\hat\lambda}{2} [k_\mathrm{B}^2 (T^2_{\Phi}-T^2_C)+4\Phi_0^2]\Phi_0\phi=0 \label{eq:Phiper}
\end{eqnarray}
In Figure \ref{fig1} 
we show the behavior of the potential for different temperatures, we see that the SB takes place at T=T$_C$. 

In the SFDM model the initial fluctuations come from inflation as in the standard CDM paradigm, 
later on the field decouples from the rest of the matter and goes through a SB which can increase the fluctuations amplitude 
forming the initial structures of the universe.

As one of the main interest of this work is finding an exact analytical solution to equation (\ref{eq:Phiper}),
we will not pursue here the task of solving it numerically. However, the numerial work done in \cite{mag12b} has 
confirm that the behavior of the SF perturbation just after the SB is what we had expected from our analysis of equation (\ref{eq:KG}).
They have analysed with some detail the evolution of a perturbation with wavelength 2Mpc and density contrast $\delta=1 \times 10^{-7}$ 
after the SB, they took as initial condition $a$=10$^{-6}$ and evolve it until $a$=10$^{-3}$. They also analysed 
the case T$\sim$T$_{C}$ and show that as the temperature decreases and goes below T$_C$, $\Phi_0$ falls rapidly to a new 
minimum where it will remain oscillating. Meanwhile, the SF fluctuation grows quickly as $\Phi_0$ approaches the 
new minimum. It is only before the SB when the SF remains homogeneous. 

From here we see that in the SFDM model, the primordial DM halos form almost at the same time due to the phase transition produced
by the SB. 
In the non-linear regime these halos can merge with other halos forming larger structures just like in the CDM model. 
Therefore, one of the main differences with the CDM model lies in the initial formation of the DM halos, they are formed very rapidly and almost at 
the same time, from here we expect that they possess similar features. 
This difference between the SFDM and CDM models can be tested by observing well formed high-redshift 
galaxies and also by comparing characteristic parameters of several DM dominated systems, for instance, by observing 
that indeed, dwarf or LSB galaxies possesses cores even at high-redshift, especially since CDM simulations of dwarf galaxies by 
\cite{gov12} suggest that their DM density profiles were initially cuspy but later on turn into core profiles due to feedback processes.

\subsection{Newtonian limit}

In this work we are interested in the galactic scale DM halos after their formation. Thus, we constrain ourselves to solve the Newtonian limit of
equation (\ref{eq:Phiper}) when $\Phi$ is near the minimum of the potential and after the SB, where we expect it to be stable.
In this limit, we also expect the gravitational potential to be locally very homogeneous in the beginning of its collapse, thus, $\dot \phi\approx 0$.
The stability of these halos will be shown elsewhere. For clarity, from here on we stop using units in which $c$=1. 
We now find an exact spherically symmetric solution for the SF when it is near the minimum of the potential ($V'(\Phi)|_{\Phi_0}\approx0$), i.e., when
$\Phi^2_0$=$\Phi^2_{\mathrm{min}}$=$k^2_\mathrm{B}($T$_C^2-$T$^2_{\Phi})/4$ and T$_{\Phi}$$<$T$_C$. 
For the Newtonian case $H =0$, $\dot \Phi_ {0} \approx 0$ and equation (\ref{eq:Phiper}) reads
\begin{equation}
\ddot{\delta\Phi}-\nabla^2\delta\Phi+ \frac{\lambda k^2_\mathrm{B}}{2 \hbar^2}(T^2_C-T^2_{\Phi})\delta 
\Phi=0. \label{eq:StC}
\end{equation} 
Equation (\ref{eq:StC}) describes the evolution of the SF fluctuation which formed at T$_{\Phi}$ after the SB. 
Moreover, this equation is linear.
In general, halos can evolve with different histories depending on their environment, 
star formation, gas accretion, and other secular processes that occur in the hosted galaxy, thus, in order to study these systems in detail we will need the 
full non-linear regime. However, we can still regard equation (\ref{eq:StC}) as a first approximation for those halos of 
isolated galaxies or of galaxies that are not heavily influenced by their neighbors, for example, 
isolated galaxies or galaxies that are in low mass density regions can be well described by this equation. Mainly, because
their halo profiles are not expected to be affected substantially from their initial profiles during their evolution.
Furthermore, we expect that the galaxies remain in the Newtonian gravitational regime, thus we can consider that the linear 
approximation is good enough for describing a galaxy.
Therefore, we interpret a solution of this system as the temperature-corrected density profile of early DM halos.

We find that the ansatz 
\begin{equation}
 \delta \Phi = \delta \Phi_0 \frac{\sin(k r)}{kr} \cos(\omega t) \label{ansatz}
\end{equation}
is an exact solution to equation (\ref{eq:StC}) provided 
\begin{equation}
 \omega^2 = k^2 c^2  + \frac{\lambda k^2_\mathrm{B}}{2 \hbar^2}(T^2_C-T^2_{\Phi}) . \label{omega}
\end{equation}
Here $\delta \Phi_0$ is the amplitude of the fluctuation. From equation (\ref{omega}) we notice that now $k=k($T$_{\Phi})$.
For an easier comparison with observations we use the standard definition of number density 
$n(\mathbf{\mathit{x}},t)$=$\kappa (\delta \Phi)^2 $, where $\kappa$ is a constant that  
gives us the necessary units so that we can interpret $n(\mathbf{\mathit{x}},t)$ as the number density of DM particles, 
as $\Phi$ has energy units.
With this in mind, we can define an effective mass density of the SF fluctuation by $\rho = m n$ and a central density by 
$\rho_{0}=m \kappa (\delta \Phi_0)^2$. It is important to note that, while $\delta \Phi_0$ is not be obtained directly from observations, 
the value of $\rho_{0}$ is a direct consequence of the RC fit, for this reason, it is preferable to work with $\rho_{0}$ instead 
of $\delta \Phi_0$.

Combining equation (\ref{ansatz}) and the definition of $n$ we obtain a finite temperature density profile
\begin{equation}
\rho (r) = \rho_{0} \frac{\sin^{2} (kr)}{(kr)^{2}},  \label{density}
\end{equation}
provided 
\begin{eqnarray}
k^2_\mathrm{B} T^2_{\Phi}& =& k^2_\mathrm{B} T^2_C - 4\Phi_0^2,  \label{dispersion} \\ 
\Phi^2_0 &=& \Phi^2_{\mathrm{min}}
\end{eqnarray}
Here $k($T$_{\Phi})$ and $\rho_{0}$=$\rho_{0}$(T$_{\Phi}$) are fitting parameters while $\lambda$, T$_C$(or $m$), and $\kappa$ are free 
parameters to be constrained by observations. 

For galaxies the Newtonian approximation gives a good description, therefore, from equation (\ref{density}) we obtain the 
mass and rotation curve velocity profiles given by
\begin{mathletters}
\begin{eqnarray}
M(r) &=& \frac{4 \pi G \rho_0}{k^2} \frac{r}{2} \biggl(1-\frac{\sin(2 k r)}{2 k r} \biggr), \\
V^2(r) &=& \frac{4 \pi G \rho_0}{2 k^2} \biggl(1-\frac{\sin(2 k r)}{2 k r} \biggr) \label{vel}. 
\end{eqnarray}
\end{mathletters}
respectively.
For comparison, we write the Einasto rotation curve profile which lately seems to give a better description of DM halos in CDM 
simulations \citep{nav10,mer06,gra06},
\begin{equation}\label{einasto}
V^2_\mathrm{E} = 4 \pi G \rho_{-2} \frac{r^3_{-2}}{r} \biggl[ \frac{e^{2/ \alpha}}{\alpha} \biggl(\frac{\alpha}{2}\biggr)^{(3/\alpha)} 
\gamma(\frac{3}{\alpha},x ') \biggr], 
\end{equation}
with $\gamma$ the incomplete gamma function given by 
\begin{displaymath}
 \gamma(\frac{3}{\alpha},x')= \int^{x'}_0 e^{-\tau} \ \tau^{(3/\alpha)-1} d\tau \nonumber
\end{displaymath}
with $x':= \frac{2}{\alpha}(\frac{r}{r_{-2}})^{\alpha}$, $r_{-2}$ is the radius in which the logarithmic slope of the density 
is $-2$, $\rho_{-2}$ is the density at the radius $r_{-2}$, and $\alpha$ is a parameter that describes the degree of curvature 
of the profile\citep{mer06,gra06}.

We now define the radius R of the SFDM distribution by the condition $\rho($R$)=0$. This fixes the relation 
\begin{equation}
 k_j \mathrm{R} = j  \pi , \ \ \  j=1,2,3,...  \label{k}
\end{equation}
where $j$ is the number of the exited state required to fit a galaxy RC up to the last measured point.
From equations (\ref{eq:StC}),(\ref{ansatz}), and (\ref{k}) we find that the halo allows exited states, i.e., the exited states are 
also solutions of equation (\ref{eq:StC}). As this equation is linear, there can be halos in combination of exited states, 
for these cases the total density $\rho_\mathrm{tot}$ is the sum of the densities in the different states, given by  
\begin{equation}
 \rho_\mathrm{tot}=\sum_j \rho^j_0 \frac{\sin^{2} (j \pi r/\mathrm{R})}{(j \pi r/\mathrm{R})^{2}},  \label{densitytotal}
\end{equation}
with $\rho^j_0$ the central density of the state $j$. It is important to note that there is only one halo formation temperature, T$_\Phi$,
also, the number of states that compose one DM halo depends of each halo, one reason is that the initial conditions change form one to another. 
Finally, we note that for each state $j$ there exists both, $\omega_j$ and $k_j$, which satisfy their corresponding equation (\ref{omega}).

An additional feature of equation (\ref{density}) is the presence of ``wiggles,'' these oscillations characteristic of SF configurations 
in exited states were also seen in \cite{sin94}. Also, we define the distance where the first peak (maximum) in the RC is reached 
as $r^1_\mathrm{max}$, this determines the first local maximum of the RC velocity, which can be obtained form equation (\ref{eq:rmax})
\begin{equation}
\frac{\cos(2 \pi j y)}{2(\pi j)^2 y} \big[ \frac{\tan(2 \pi j y)}{2 \pi j y} -1\big] = 0, \label{eq:rmax}
\end{equation}
where we used equation (\ref{k}) and $y:=(r^1_\mathrm{max}/\mathrm{R})$. 

\section{DISCUSSION}

We have seen that within our scenario of dark matter and galaxy formation the DM halos are naturally cored, i.e., their 
central density profiles are finite and do not diverge, on the contrary, they behave as $\rho$$\sim$$r^0$ for small $r$, it
is important to note that the core is obtained from the model and not assumed, thus, this is an alternative way to 
solve the cusp/core discrepancy without adding large amounts of baryons in simulations\citep{kuz11b}.

In the SFDM model the big DM halos (density fluctuations) form after the SB and grow only after the field rolls to the minimum
of the potential, same which varies with the temperature. Nevertheless, during this time the halos are not in thermal equilibrium, 
locally the temperature is different from place to place. Therefore, the initial size of the condensation depends on the local halo temperature. 
From equations (\ref{omega}) and (\ref{k}) we see that the size of the DM configuration, specified by R, is now temperature-dependent, therefore, 
as in \cite{har11b}, we also solve the problem of having a unique scale lenght for all halos, but now in a new way, by using
the SB mechanism.
Therefore, different formation temperatures of galactic halos may result in different DM halo sizes.

\begin{figure}
\begin{tabular}{cc}
\includegraphics{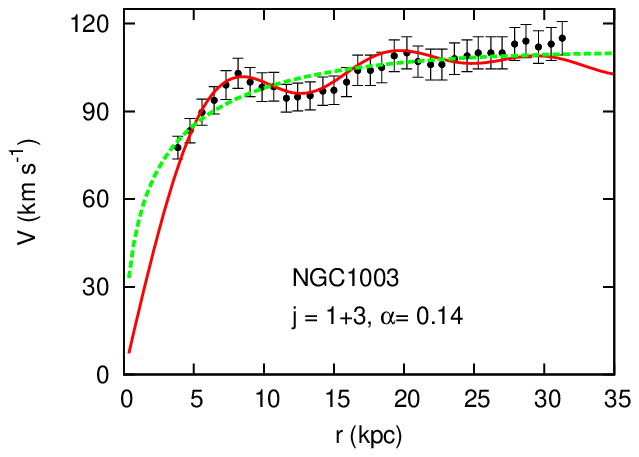} \\
\includegraphics{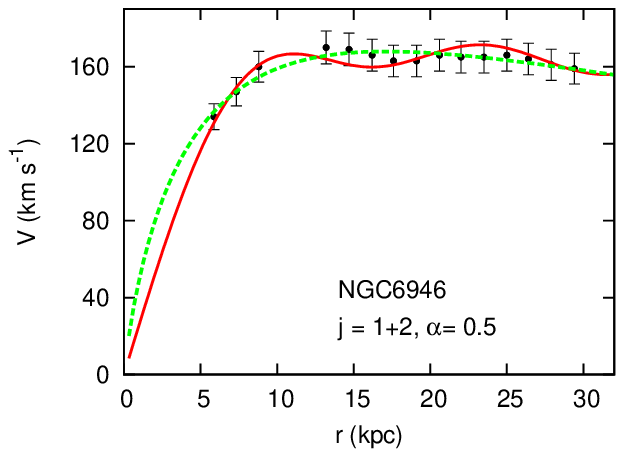} \\
\includegraphics{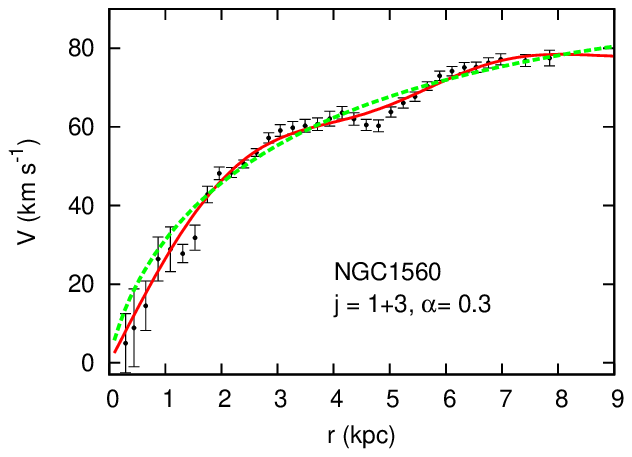}
   \end{tabular}
\caption{Rotation curve fits to three LSB galaxies. \textit{top panel}: NGC 1003, \textit{middle panel}: NGC 6946, \textit{bottom panel}: NGC 1560.
Solid lines (red in the online version) are the fits using the SFDM model, dashed line (green in the online version) represents Einasto's fits, and
triangles are the observational data. In NGC 1560 we see that the dip at $r$$\approx$5kpc is reproduced more accurately in the 
SFDM profile. Einasto fits show different values of $\alpha$ in each galaxy, suggesting a non-universality in the DM halos, the 
same is concluded in the SFDM model.}
\label{fig:2}
\end{figure}

In Figure \ref{fig:2} 
we show the RC fits of three galaxies using the minimum disk hypothesis (neglecting the baryonic component)
taken from a high-precision subsample of \cite{mcg05} combined with \cite{bro92} for NGC 1560. 
The sample given in \cite{mcg05} consists of resolved rotation velocity measurements, with accuracy of 
5\% or better. The error bars in NGC1560 are taken from the data in \cite{bro92} and we give uniform weight of 5\% to 
the other two galaxies.
We compare equation (\ref{vel}) (solid line) with equation (\ref{einasto}) (dashed line) and notice that these galaxies 
present two features, long flat tails in the outer region and wiggles, though the latter are still debated and a much larger sample 
is needed to verify their existence in RCs. However, in our sample, they do seem to be present, even after overstimating the error bars. 
The wiggles (small oscillations) are perfectly reproduced by the SFDM model by using combinations of exited states, the value of $j$ 
that appears in the panels of Figure 2 specifies the required combination of states for the fit shown. It is important to highlight the SFDM fit of 
NGC 1560, we note it is as accurate as the one displayed in MOND models \citep{gen10,san09}. 
This combination of states in our RC fits suggests that there is not a universal DM profile, 
some reasons could be that 1) the subsequent evolution determines the final profile, as happens with CDM halos, 2) a collision of 
two halos with different states formed a halo with the combination of states that we observe today, and 3) the halo formed with the currently 
observed states and has remained unaltered for a long time. Further research is necessary to determine the most likely explanation.

As for the observed long flat tails, we have that the flat outermost region is a consequence of considering exited states, 
same behavior that was present in previous works \citep{sin94,ber10} which used T=0. However, the main difference now is that by 
considering T$\neq$0 we can accomodate exited states in halos and expect them to be stable due to the DM thermal and repulsive self 
interactions. 
Moreover, by considering for the first time both finite DM temperature corrections and exited states, we 
obtained an excellent agreement with the observed data (see Figure \ref{fig:2}), suggesting that large 
quantities of baryons are not essential to fit RCs in our model. This precludes the necessity of including 
large amounts of gas, stars, and feedback processes to flatten the inner regions of RCs. 
Due to the accurate fits obtained, we expect that adding only the observed amount of baryons will be enough 
to reach perfect agreement with RC observations and the observed luminosity. However, this will be verified in a future work.

\begin{figure}
\flushleft
\plotone{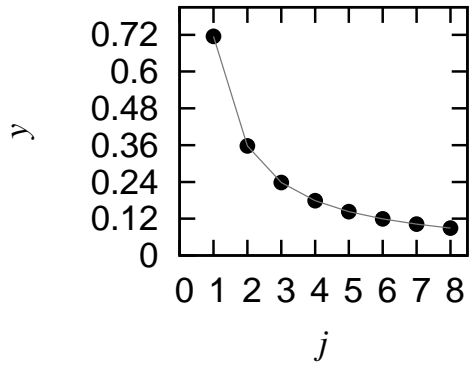}
\caption{We plot the relation between $y$ and $j$ obtained by solving equation (\ref{eq:rmax}).
Notice that for halos with large exited states (large $j$), the first maximum is attained at a smaller $y$ for a fixed radius.}
\label{fig:3}
\end{figure}

Regarding the second problem about fitting the maximum of the RC and the outer regions at the same time,
we notice from Figure \ref{fig:2} that we do not have such problem anymore, in contrast to previous works \citep{har11,har11b,rob12},
the difference is mainly due to the combination of states in halos. We can estimate a lower bound for $j$, the minimum  
exited state necessary to agree with the data up to the last measured point, with the following rule of thumb: we find the 
value of $y$ by identifying from the RC the last measured point as R and the first maximum (first visible peak) as 
$r^1_\mathrm{max}$, then, we look for the closest value of $j$ associated with this $y$ in Figure \ref{fig:3} (the values of Figure \ref{fig:3}
are determined by equation (\ref{eq:rmax})),
this provides us with the dominant state in the center of the galaxy and at the same time 
with the minimum state required to fit current observations. 
It is a lower bound because upcoming observations might go beyond R, in that case R will increase 
and $y$ will decrease, implying larger values of $j$. 

As some final remarks, we see from Figure \ref{fig:2} that Einasto fits are in general agreement in the outer regions, 
but the wiggles cannot be reproduced with DM only. In fact, if we want to reproduced the oscillations seen 
in high-resolution RCs with a non-oscillatory DM profile (NFW, Einasto's, Burkert's etc.),
we must include the gas and stars dynamics in the simulations\citep{fam12}, it would be interesting to show the stability 
of the oscillations after including baryons, as this might be a challenging task in LSBs galaxies due to their 
low gas surface density.

In addition to the fact that the fits to the observations made with the Einasto profile are good, except for the wiggles and 
possibly the central region, we notice that the parameter $\alpha$ changes for each galaxy. As noted in previous 
works \citep{mer06,nav10,jin00,gen07}, the change in $\alpha$ implies that halos do not possess a universal profile, i.e., 
if we take Einasto profile to be the best representation of the simulated DM halos in a CDM environment, 
then we should expect to see a non-universality in the halos of galaxies, this is exactly the same result we have 
obtained directly from the SFDM model but without assuming a priori a DM density profile.

\section{CONCLUSIONS}

In this paper we gave a physically motivated extention to the SFDM model that includes the DM temperature corrections to the first loop 
in perturbations.  
We propose a $Z_2$ spontaneous symmetry break of a real scalar field as a new mechanism in which the early DM halos form. 
When the real scalar field rolls down to the minimum of the potential, the perturbations of the field can form and 
grow.
We give an exact analytic solution for an static SF configuration, which in the SFDM model represents 
a DM halo. This solution naturally presents a flat central density profile, it can accomodate more than just the ground state
as now the temperature T$\neq$0, and solves previous discrepancies in rotation curve fits at T=0, for instance,
having a constant halo radius for all galaxies and, in the case of large galaxies, the incapability to fit at the same time 
the inner and outermost regions of RCs. 

Additionaly to solving these two disagreements we mention why it does not seem necessary to include high amounts of feedback to 
fit and reproduced the inner core and ``wiggles'' found in high-resolution RCs. Also, the model can be tested with 
high redshift observations, the SF model predicts initial core profiles as opposed to the initially cuspy ones found in CDM simulation 
which are expected to flatten due to redistribution of DM by astrophysical processes. 
Finally, both the Einasto and SFDM RC fits suggest the non-universality of the DM halos, though the latter claim is still not 
final and further work has to be done in such direction. We strongly believe that exploring further the SFDM looks promising to 
unravel the mystery of dark matter.

\acknowledgments

This work was supported in part by DGAPA-UNAM grant IN115311 and by CONACyT Mexico grant 49865-E.


\end{document}